\documentclass[a4paper, amsfonts, amssymb, amsmath, preprint, showkeys, nofootinbib, twoside]{revtex4-1}
\usepackage[english]{babel}
\usepackage[utf8]{inputenc}
\usepackage{float}
\usepackage[colorinlistoftodos, color=green!40, prependcaption]{todonotes}
\usepackage{amsthm}
\usepackage{mathtools}
\usepackage{physics}
\usepackage{xcolor}
\usepackage{graphicx}
\usepackage[left=23mm,right=13mm,top=35mm,columnsep=15pt]{geometry} 
\usepackage{adjustbox}
\usepackage{placeins}
\usepackage[T1]{fontenc}
\usepackage{lipsum}
\usepackage{csquotes}
\usepackage[pdftex, pdftitle={Article}, pdfauthor={Author}]{hyperref} 
\usepackage{subfigure} 

\begin{document}
\title{Black Hole Explosions as Probes of New Physics}

\author{Kevin Federico}
\email{kcf225@lehigh.edu}
\affiliation{Department of Physics, Lehigh University, Bethlehem, PA, 18015, USA}
\affiliation{Santa Cruz Institute for Particle Physics (SCIPP),
Santa Cruz, CA 95064, USA}

\author{Stefano Profumo}
    \email[Correspondence email address: ]{profumo@ucsc.edu}
    \affiliation{Department of Physics, University of California, Santa Cruz (UCSC),
Santa Cruz, CA 95064, USA}
\affiliation{Santa Cruz Institute for Particle Physics (SCIPP),
Santa Cruz, CA 95064, USA}

\date{\today} 

\begin{abstract}
\noindent The final stage of black hole evaporation is a potent probe of physics beyond the Standard Model: Hawking-Bekenstein radiation may be affected by quantum gravity ``memory burden effects'', or by the presence of ``dark'', beyond-the-Standard-Model degrees of freedom in ways that are testable with high-energy gamma-ray observations. We argue that information on either scenario can best be inferred from measurements of the evaporation's lightcurve and by correlating observations at complementary energies. We offer several new analytical insights in how such observations map on the fundamental properties of the evaporating black holes and of the possible exotic particles they can evaporate into.
\end{abstract}


\maketitle

\section{Introduction}
Evaporating black holes are excellent probes into the otherwise unreachable domains of very high energy physics and even into the structure of quantum gravity (see e.g. \cite{Baker_2022, Baker_2023, Boluna:2023jlo}). Ever since Hawking's seminal paper \cite{Hawking:1974rv}, it has been firmly established that black holes can be approximately regarded as thermal bodies with a temperature inversely proportional to their mass --- thus evaporating by emitting particle radiation and  losing mass as a result. One would expect that at the end of their ``life'', when the temperature gets increasingly close to the Planck scale, they should evaporate into any particle of any mass charged under any interaction, including  additional, beyond-the-Standard-Model degrees of freedom possibly entirely secluded. However, all of the familiar stellar-mass and supermassive black holes have a temperature much lower than, for instance, the cosmic microwave background, thus effectively absorbing particles and gaining mass rather than losing it. Thus, if one were to use black holes as a probe into the particle spectrum, the focus must be on black holes with much smaller masses, and thus of non-standard, or primordial, rather than stellar origin \cite{Carr:2021bzv}. A simple calculation  shows that a black hole expiring today (i.e. a black hole with a lifetime approximately equal to the age of the universe) would need to have a mass in the early universe on the order of $M_{\rm BH} \sim 5\times10^{14} $ g \cite{MacGibbon_2008}. Such light black holes may originate at early times, with a large-enough mass for them to be evaporating now, or even emerge in the late universe in some scenarios \cite{Carr:2021bzv}.

In principle, a black hole radiates any quantum degree of freedom with an associated mass less or around its temperature, with heavier particle emission exponentially suppressed by the usual Boltzmann factor. In the standard picture of black hole radiation, the black hole will continue to lose mass, and thus increase temperature, until they both approach the Planck scale.  Thus, evaporating primordial black holes will radiate the entire sub-Planckian particle spectrum, making them unique probes of very high energy physics. Previous studies have investigated in part the observational signals of such events \cite{Ukwatta_2016_Analytics,Ukwatta_2016_IPNGRB,Baker_2022, Boluna:2023jlo}. However, the standard evaporation picture, as investigated by the authors above so far, is not the only possible one. 

There is strong evidence that the Standard Model is incomplete, yet even  the most powerful current accelerators --- reaching energies up to a few TeV --- provided, so far, little-to-no evidence or guidance on physics beyond the Standard Model.  Observation of an evaporating, light black hole could provide a wealth of information about the particles and degrees of freedom above this energy scale \cite{Baker_2023} and thus on physics at mass scales we may never access with accelerators. Furthermore, black hole spectra would also be excellent tools for particle dark matter searches \cite{Baker_2022}. 

The present study is concerned with ways in which the  final ``explosion'' (as Hawking referred to as \cite{Hawking:1974rv}) may differ from the standard picture, and the concrete possibility of inferring, via ground and space-based gamma-ray telescopes, the mass-scale and size of dark sectors and/or the features of other scenarios that may affect the Hawking-Bekenstein evaporation process. We will examine here searches for additional degrees of freedom (which may originate from a broad range of new physics hypothetical scenarios such as dark sectors \cite{Agrawal:2021dbo}, SUSY \cite{bilal2001introductionsupersymmetry}, and NNaturalness\cite{Arkani_Hamed_2016}, to name a few). In addition, we will also look at how a black hole's radiation and corresponding spectra may be altered by the so-called memory burden effect \cite{dvali_2024_memoryburdeneffectblack, Dvali_2020, alexandre_2024}. To study in details these various scenarios we will primarily be utilizing the \texttt{BlackHawk} code \cite{BlackHawk1.0,BlackHawk2.0}, which enables us to reliably calculate the detailed spectra and lightcurves of the evaporating black holes. 

The remainder of this study is structured as follows: in the next section we discuss the general theoretical framework; 
the following Sec.~\ref{sec:observatories} discusses the form of the observational signals and the specific telescopes we choose to consider in this study;
Sec.~\ref{sec:dof} and \ref{sec:memoryburden} discuss in detail the cases of additional degrees of freedom and memory burden effects, respectively; Sec.~\ref{sec:results} presents our results, and the concluding Sec.~\ref{sec:discussion} contains our discussion and conclusions.

\section{Theoretical Framework}\label{sec:theory}

The temperature of a Schwarzschild black hole (which we hereafter assume -- albeit charged and rotating black holes' evaporation differs slightly in the lightcurve and spectra; also notice that hereafter we will work in natural units, $c = \hbar = \kappa_b = G = 1$)  is
\begin{align}\label{temp}
T = \frac{1}{8\pi M}.
\end{align}
Standard black-body theory, in the background of the Schwarzschild metric,  implies that the emission rate of particle $i$ from a black hole of mass $M$, and corresponding temperature $T$, reads:
\begin{align} \label{emission_rate_eq}
\frac{d^2N^{i}}{dtdE} &\simeq  \frac{{n}^i\ \Gamma^i(M,E)}{2\pi(e^{E/T}-(-1)^{2s_i})}.
\end{align}
In the expression above, $n$ refers to the internal number of degrees of freedom of  particle $i$, $s$ is the particle's spin, $E$ is the energy of the particle radiated, and $\Gamma^i(M,E)$ is the corresponding greybody factor \cite{Page_1976_I}, which encodes the probability that a particle escapes to spatial infinity in the background of the Schwarzschild metric \cite{Page_1976_I, Page_1976_II, Page_1976_III}. The computation of the greybody factors is analytically and numerically complex; here, we use the values for $\Gamma^i(M,E)$ accurately computed and tabulated by the \texttt{BlackHawk} code \cite{BlackHawk2.0}. 

It is useful  to discuss the emission rate in terms of the dimensionless parameter $x = E/T$, with $T$ the hole's temperature, Eq.~\eqref{temp}. 
 As in \cite{Ukwatta_2016_Analytics}, we find that the emission rate in Eq.~\eqref{emission_rate_eq} generally peaks at $x \sim 6$. We have confirmed this fact fully numerically and in detail with \texttt{BlackHawk}. We can then relate the temperature to the remaining lifetime and solve for $x \sim 6$ to obtain when the peak photon emission occurs in the standard case, and, neglecting the finite energy range  which observatories are limited to (to be discussed below), this to a peak in the remaining lifetime $\tau$, as a function of energy $E$, and a corresponding peak temperature $T_{\rm peak}$ given by \cite{Ukwatta_2016_Analytics}
\begin{align}\label{timevtemp}
\tau_{peak} \simeq\left(\frac{4.68\times10^4}{E}\right)^3\text{s}, \qquad T_{\rm peak} \simeq 7.8\times 10^3\left(\frac{\tau_{\rm peak}}{1\text{s}}\right)^{-1/3} \text{GeV}
\end{align}
Note that there are two different classes of radiated photons  \cite{Coogan:2019qpu, Coogan:2020tuf, Coogan:2022cdd, Koivu:2024gjl}: The first are {\it primary} photons, i.e. photons promptly produced in the evaporation process; The second kind are  {\it secondary} photons, radiated for instance in bremsstrahlung processes associated with charged particles, or resulting from the decay of hadronized strongly-interacting evaporation products. 
The computation of the primary spectra --- as seen above --- is fairly straight-forward. Calculating the contributions of the secondary spectra however is much more involved. For this, we use the HAZMA hadronization code \cite{hazma} at low energy and PYTHIA \cite{pythia} for higher energies.

As the black hole radiates, it loses mass. As shown in detail in Ref.~\cite{Page_1976_I,Page_1976_II,Page_1976_III} the mass changes as:
\begin{equation} \label{dmdt}
\frac{dM}{dt} = -\frac{\alpha(M)}{M^2},\quad {\rm where} \quad  
\alpha(M) = M^2\sum_i\int_0^\infty\frac{d^2N^{i}}{dtdE}\,E\,dE.
\end{equation}

As more degrees off freedom are added to the model, $\alpha(M)$ --- known as the {\it Page factor} --- and thus $\frac{dM}{dt}$, will change. Even if new degrees of freedom are not coupled to the electromagnetic field and thus producing photons, they cause a significant difference in the observable light curve \cite{Baker_2022, Boluna:2023jlo}, speeding up the overall evaporation process, as we discuss in detail below.

The main concern of this paper is to compute  emission rates for different non-standard scenarios corresponding to the size --- i.e. number of degrees of freedom --- and mass-scale of dark sectors. Observationally, the  quantity we are concerned with is $\frac{dN_\gamma(t)}{dt}$, the rate of photon emission as a function of time, albeit other evaporation products may be critical as well (see e.g. \cite{Korwar:2024ofe}). This flux corresponds to the integral of Eq.~\eqref{emission_rate_eq}, namely
\begin{align}
\frac{dN_\gamma}{dt} &= \int_{E_{\rm min}}^{E_{\rm max}}\frac{d^2N^{i}}{dtdE}\,dE,
\end{align}
where $E_{\rm min}$ and $E_{\rm max}$ represent the effective low- and high-energy bounds of the telescope under consideration\footnote{Of course such bounds are not abrupt, but, rather, correspond to the energy where the telescope's effective area drastically decreases.}. In our case, this will be fixed, effectively, by the specifications of the detectors under consideration.


\section{Observatories}\label{sec:observatories}

We focus here on two different detectors,  with energy ranges comprising a high energy band --- approximately 100 GeV - $10^5$ GeV (HAWC) and  a low energy band --- approximately $0.1$ GeV - 300 GeV (Fermi-LAT). 

For the first energy band, we focus on a schematic approximation of the High-Altitude Water Cherenkov Observatory (HAWC)  on the Sierra Negra volcano in Mexico \cite{HAWC_first_results}. Sitting at an altitude of 4100 meters, HAWC consists of 300 steel tanks of water. As high-energy gamma rays impact the atmosphere, they produce showers of charged particles. These charge particles pass into the tanks and emit Cherenkov radiation, picked up by  photomultiplier tubes. Given HAWC's field of view of about 2 steradians and a maximal effective area of about $10^5\, \text{m}^2$\cite{MEGA_HAWC}, this is an ideal tool to search for exploding black holes -- as done e.g. in Ref.~\cite{HAWC_first_results}. HAWC data and its projected sensitivity have been used before in similar studies investigating black hole explosions \cite{Baker_2022, Boluna:2023jlo}. 


The second detector we consider is the Fermi Large Area Telescope (LAT) \cite{Fermi-LAT:2009ihh}: Fermi-LAT is a space-based gamma-ray telescope operating since 2008, with an approximate sensitivity over the energy band of interest here of $3 \times 10^{-9}$ photons cm$^{-2}$ s$^{-1}$, 
making it a promising tool for detecting black hole explosions; unlike HAWC, however, the relatively small field of view implies that the observatory must be pointed at a specific section of sky in order to observe the explosion.  

As the photons travel from the source to the detector, they spread out radially. To count the total number of photons collected from a black hole explosion event by a given observatory, one integrates Eq.~\eqref{emission_rate_eq} while taking into account the radial flux depletion and the energy-dependent effective area as
\begin{align}
N_{obs} &= \int_{t_0}^{t_f}\int_{E_0}^{E_f}\frac{A_{\rm eff}(E)}{4\pi R^2}\frac{d^2N}{dtdE}\,dE\,dt,
\end{align}
where $A_{\rm eff}$ corresponds to the effective area of the detector in the relevant energy range and $R$ the distance to the exploding black hole. Each detector has a different effective area that determines how sensitive they are to a given flux of photons, which take into account numerically and take from the quoted detector's performance \cite{Fermi-LAT:2009ihh, FermiLAT, HAWC_first_results, MEGA_HAWC}. 

\section{Additional Degrees of Freedom}\label{sec:dof}
Recall from Eq. \eqref{emission_rate_eq} that as one adds additional degrees of freedom, the black hole emission rate changes. Looking beyond the standard model, we consider here, in a model-independent way, scenarios with additional degrees of freedom at a common mass scale $M$ and in a number $n$, and study how a dark sector defined by ($M,n$)  affects the standard black hole evaporation picture and its observational features. We assume there is negligible coupling between the extra degrees of freedom and the standard model --- thus effectively treating the new degrees of freedom as ``dark'', i.e. not emitting photons.

The most obvious consequence of the presence of additional degrees of freedom is that the emission rate, and thus the lifetime and mass loss rate of the black hole, will change. Modifying the \texttt{BlackHawk} code to work with additional degrees of freedom, we find that the Page factor $\alpha(M_{\rm BH}$ (we utilize hereafter the subscript BH to distinguish the black hole mass from the mass of the dark degrees of freedom)  diverges from the standard emission picture when the temperature of the black hole becomes greater than the mass scale of the additional degrees of freedom $M$, as expected, and by different amounts depending on the number of such additional degrees of freedom. We show quantitatively how below. The size of the deviation from the standard case is, in turn, dependent on the number degrees of freedom added,  $n$. 

\begin{figure}[t]
    \includegraphics[width=0.75
    \textwidth]{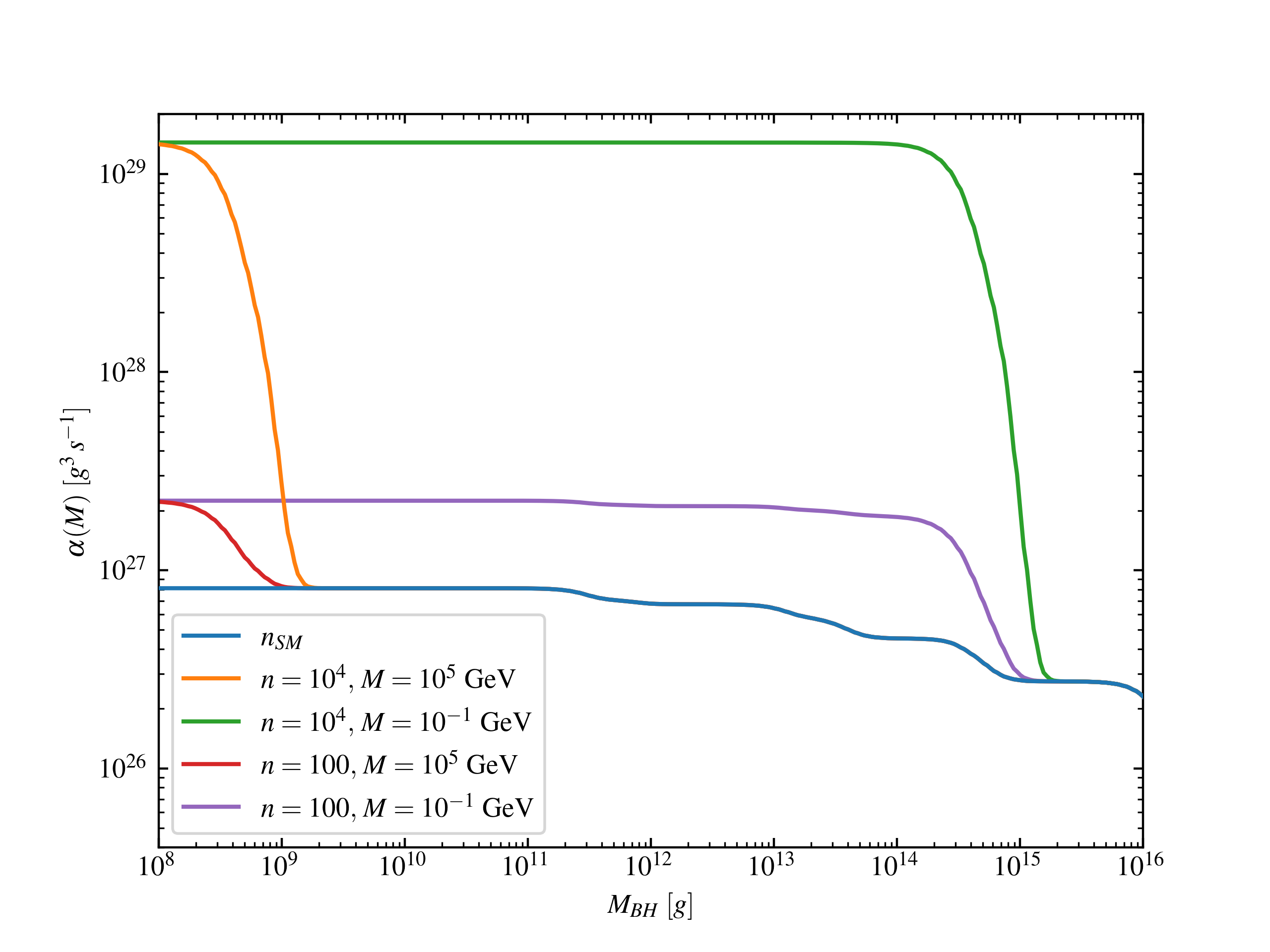}
    \caption{The page factor $\alpha(M_{\rm BH})$, for the standard case ($n_{\rm SM}$) and for a variety of choices of number $n$ and mass scale $M$  of additional degrees of freedom.}
    \label{fig: alpha plot}
\end{figure}

We show the Page factor explicitly in Fig.~\ref{fig: alpha plot}, as a function of the black hole mass for different choices of $n$ and $M$, compared to the case with no additional degrees of freedom (blue line). The figure reflects exactly what is expected from classical black hole thermodynamics \cite{Hawking:1974rv, Page_1976_I}, albeit with great numerical precision. 

\begin{table}[t]
    \centering
    \begin{tabular}{|c|c|c|c|} \hline 
         $M$ (GeV)&  $n$&  $\tau_{\rm crit}$ Eq.~\eqref{critical time} (s) & $\tau_{\rm crit}$ \texttt{BlackHawk} (s)\\ \hline 
         $10^5$&  $10^2$& 0.71 & $0.55$\\ \hline 
         $10^5$&  $10^3$& 2.40 & $2.66$\\ \hline 
         $10^5$&  $10^4$& 5.68 & $5.53$\\ \hline 
         $10^5$&  $10^5$& 11.09 & $11.47$\\ \hline 
         $10^5$&  $10^7$& 30.43 & $27.60$\\ \hline \hline
         $10^{-1}$&  $10^2$& $7.10 \times 10^{17}$ & $6.47\times 10^{17}$\\ \hline 
         $10^{-1}$&  $10^3$&  $2.40 \times 10^{18}$& $2.72 \times 10^{18}$\\ \hline 
         $10^{-1}$&  $10^4$& $5.68 \times 10^{18}$ &$5.73 \times 10^{18}$\\ \hline
         $10^{-1}$&  $10^5$& $1.11 \times 10^{19}$ &$1.10 \times 10^{19}$\\ \hline
         $10^{-1}$&  $10^7$& $3.04 \times 10^{19}$ &$2.79 \times 10^{19}$\\ \hline\hline
 $10^{-4}$& $10^2$& $7.10 \times 10^{26}$&$1.80 \times 10^{27}$\\ \hline
 $10^{-4}$& $10^3$& $2.40 \times 10^{27}$&$7.28 \times 10^{27}$\\ \hline
 $10^{-4}$& $10^4$& $5.68 \times 10^{27}$ &$1.62 \times 10^{28}$\\ \hline
    \end{tabular}
    \caption{$\tau_{\rm crit}$ values for different dark sectors ($M,n$).}
    \label{tab:tcrit}
\end{table}

We now intend to find the precise time-to-expiration corresponding to the onset of the effect of the additional degrees of freedom on the evaporation rate for a given ``dark sector'' ($M,n$). As stated above, this occurs at a time-to-expiration, which we call  $\tau_{\rm crit}$, around a mass corresponding to the temperature of the black hole becoming greater than the mass scale of the additional degrees of freedom.  Recall the temperature-time relation of Eq. \eqref{timevtemp}, which reflects the case with no additional degrees of freedom, $$T \simeq 7.8 \times 10^3\ {\rm GeV}\  \left(\tau/1\ {\rm sec}\right)^{-1/3}.$$
%
Adding a number of degrees of freedom $n$, the time-temperature relation is affected: naively, since the Page factor $\alpha\propto n+n_{\rm SM}$, and $T\sim \alpha^{-1/3}$, we expect a correction of order 
\begin{equation}\label{temperature}
T(\tau)\to T\left(\frac{n+n_{\rm SM}}{n_{\rm SM}}\right)^{1/3}.
\end{equation}
The equation above of course does not include explicitly the temperature/mass dependence of $\alpha(M_{\rm BH})$, so it is only valid at temperatures much larger than the mass scale at which the additional degrees of freedom exist. 
In this regard, notice that from Fig.~\ref{fig: alpha plot} it is evident that the time when the additional degrees of freedom set in depends  not only on the mass scale but also on the number of degrees of freedom. To quantify this trend, by modifying Eq.~\eqref{timevtemp} we find that the time when the additional degrees of freedom are radiated, and therefore when they leave a detectable imprint on the lightcurve, is, in the limit $n\gg n_{\rm SM}$, 
\begin{align}\label{critical time}
\tau_{\rm crit}(n,M) \simeq 0.07\ {\rm sec}\ \left(\ln(n)\frac{4.68\times 10^4 \, \text{GeV}}{M}\right)^{3}. 
\end{align}
The equation above is derived as follows. The logarithmic dependence on $n$ stems from the contribution to the Page factor of the additional degrees of freedom, which is approximately $\sim n\exp(-M/T)$; $\tau_{\rm crit}$ corresponds to the temperature at which that contribution is of order 1, or $T\sim M/(\ln n)$; since $T\sim \tau^{-1/3}$ the formula in \eqref{critical time} follows.

Table \ref{tab:tcrit} shows the  the value of $\tau_{\rm crit}$ as computed from Eq.~\eqref{critical time} for the mass scales we examine in this paper, as well as the value of the divergence from the $n=0$ as computed numerically from \texttt{BlackHawk} as the point where the slope in the temperature-time relation starts diverging from the standard case, for a number of choices of $n$ and $M$; the table clearly validates our discussion above and Eq.~\eqref{critical time}. 

\begin{figure}[t]
    \includegraphics[width=0.75
    \textwidth]{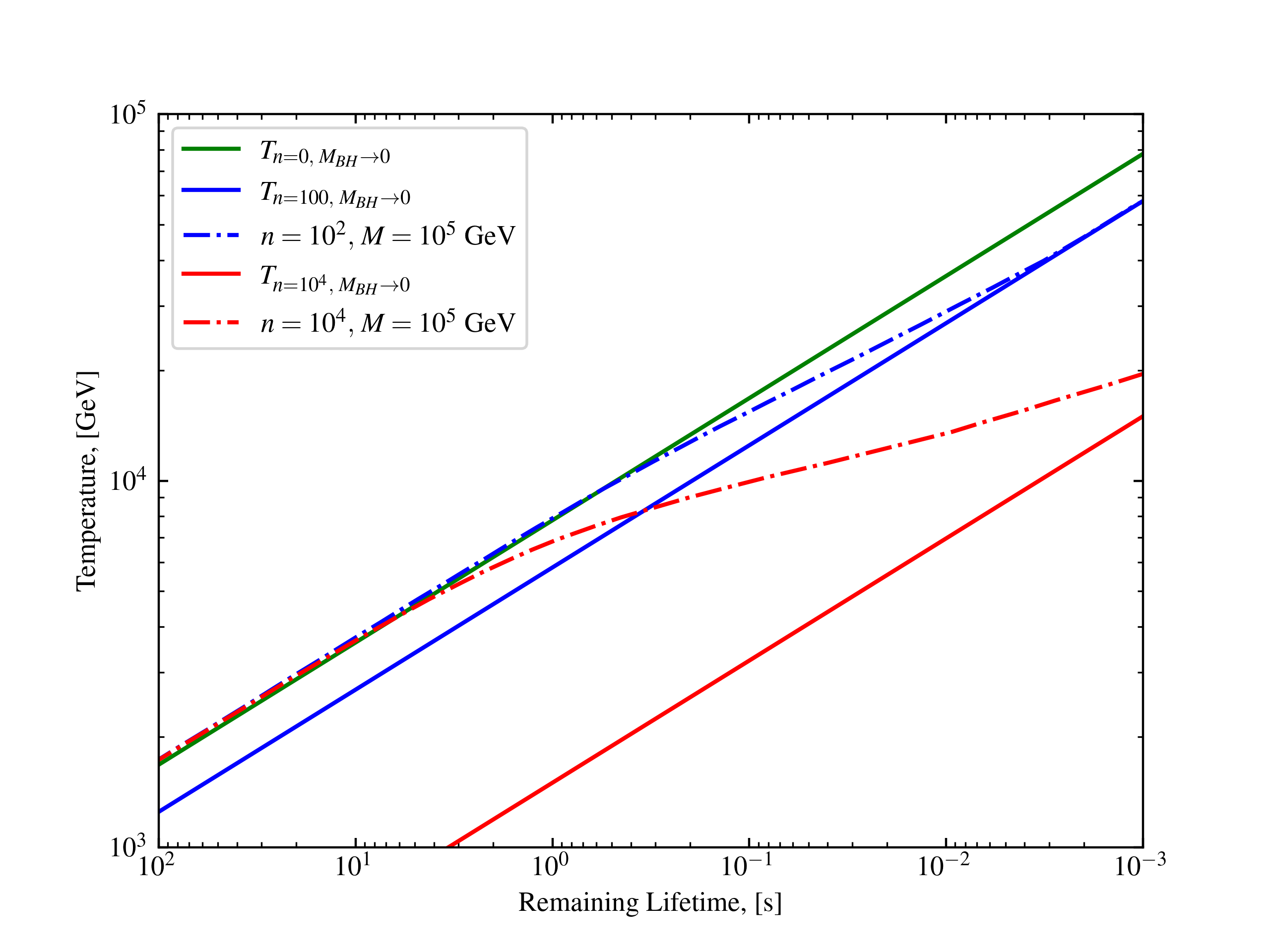}
    \caption{The relation between black hole temperature and remaining lifetime in the presence of $n$ additional degrees of freedom at a mass scale $M=10^5$ GeV; solid lines indicate the asymptotic temperature-lifetime relation in the limit $\tau,M_{\rm BH}\to 0$.}
    \label{fig:taucrit}
\end{figure}

At times later than $\tau_{\rm crit}$, the temperature evolves with respect to the remaining lifetime according to the following equation:
\begin{align}\label{timevtempmod}
T = A(n)\left(\frac{\tau}{1\text{s}}\right)^{-1/3}\times 10^3\ \text{GeV},
\end{align}
where $A(n)$ is a parameter that is related to the number additional degrees of freedom $n$ as
\begin{equation}
    A(n_{\rm SM}) = 7.8,\  A(n_{\rm SM}+10^2) = 5.8,\  {\rm and}\  A(n_{\rm SM}+10^4)= 1.5.
\end{equation} 
The factors $A$ above can be readily understood from Eq.~\eqref{temperature}, where, accounting for the suppression in the emission rate of fermions $n_{\rm SM}\simeq 70$, and
\begin{equation}
A(n)\simeq A(n_{\rm SM})\left(\frac{n_{\rm SM}}{n+n_{\rm SM}}\right)^{1/3}.
\end{equation}
We show results for the temperature-time relation, for a mass scale $M=10^5$ GeV, in Fig.~\ref{fig:taucrit}.
Note how each curve starts on the $T(n = 0)$ line, corresponding to no additional degrees of freedom, and diverges at lifetimes corresponding to temperatures close to the mass scale $M=10^5$ GeV, with the offset dictated by Eq.~\eqref{critical time}. The figure indeed confirms the behavior analytically discussed above, with higher $n$ cases diverging sooner. After the onset of the additional degrees of freedom, each curve lines up asymptotically with its corresponding $M_{\rm BH}\to 0$ asymptote.

After $\tau < \tau_{\rm crit}$ the only relevant parameter is the number of additional degrees of freedom --- the mass scale of said degrees of freedom effectively becomes irrelevant, as shown by the solid, asymptotic curves in Fig.~\ref{fig:taucrit}.

Notice that since $\tau_{\rm crit}$ is proportional to the inverse cube of $M$, models with large $n$ and small $M$ will reach $\tau_{\rm crit}$ very early in their lifetime. This means that for such models, the black hole's emission will be parameterized essentially only by the additional degrees of freedom, making the mass scale, therefore, irrelevant; in such case, the determination of the new degrees of freedom's mass scale $M$ is observationally unfeasible, unless observations start {\it prior} to $\tau_{\rm crit}$. For example, the last two mass scales considered in the above table have $\tau_{\rm crit} > \tau_{U}$, the latter being the age of the universe: in that case, the black hole is effectively {\it always} radiating the additional degrees of freedom. This means that no observer could distinguish between models with $M < 0.1$ GeV, where $\tau_{\rm crit} \sim \tau_{U}$. We discuss this in detail below.

\section{Memory Burden Effects}\label{sec:memoryburden}
The memory burden effect, introduced in a series of papers, e.g.  Ref.~\cite{dvali2018microscopicmodelholographysurvival}, is the claim, in short summary, that ``{\it information loaded into a system resists its decay}'' (see also \cite{dvali_2024_memoryburdeneffectblack}). This effect has very interesting implications for black holes, as they effectively are systems storing huge amounts of information. Memory burden would imply that as a black hole radiates it will at some point become saturated with information which will stabilize it against its own decay (again, we direct the Reader keen on further details and a more thorough discussion to \cite{dvali2018microscopicmodelholographysurvival}). This saturation and subsequent possible stabilization is expected to happen once the black hole has lost about {\it half} of its initial mass. After stabilization due to memory burden, the black hole loses mass at a significantly {\it slower} rate, thus affecting the corresponding lightcurve\footnote{SP thanks Jos\'e Ramon Espinosa for suggesting the exploration of this aspect.}.

Using the formalism in Ref.~\cite{alexandre_2024}, we  express the black hole's lifetime as a function of $n_{MB}$  powers of the entropy $S$, relating it to the black holes mass and the Planck mass as:

\begin{align}\label{lifetime_mb}
S \sim \left(\frac{M_{\rm BH}}{M_p}\right)^2, \qquad r_{\rm BH} \sim \left(\frac{M_{\rm BH}}{M_p^2}\right);\qquad  t^{n_{MB}}_{\text{lifetime}} \sim S^{1+n_{MB}}r_{\rm BH} \sim \frac{M_{\rm BH}^{3+2n_{MB}}}{M_p^{4+2n_{MB}}}.
\end{align}

%

The parameter $n_{MB}$ thus controls the relative ``strength'' of the memory burden. Naturally, $n_{MB} = 0$ corresponds to the standard emission rate, with no memory burden effect. We assume $n_{MB}$ to be a non-negative integer, but in principle one could also investigate non-integer values (this however would assume that the decay rate is not necessarily  analytic in $S$ \cite{alexandre_2024}).

We are interested here in how this modified lifetime and mass loss scaling affects the observable counterparts to the black hole's explosion. Solving for $M$ in Eq.~\eqref{lifetime_mb} when $t^{n_{MB}}_{\text{lifetime}} \sim 10^{17} $ s one  finds:

\begin{align}\label{mass_Mb}
M_{\rm BH}^{(n_{MB})} &\sim 10^{-6}\text{g}\left(\frac{10^{17}\ \text{s}}{10^{-42}\ \text{s}}\right)^{\frac{1}{3+2n_{MB}}}. 
\end{align}


Black holes burdened by memory will still follow the same mechanics of particle emission discussed in Section \ref{sec:dof}, now with a modified mass loss rate. Following \cite{alexandre_2024} one can differentiate Eq.~\eqref{lifetime_mb} to find:
\begin{align}\label{mb_mass_loss}
\frac{dM_{\rm BH}^{(n_{MB})}}{dt} \sim - M_p^2 \left(\frac{M_p}{M_{\rm BH}}\right)^{2+2n_{MB}}.
\end{align}
The expression above, for $n_{MB} = 0$ recovers the functional form of Eq.~\eqref{dmdt}, as it should.

We modified the \texttt{BlackHawk} code, specifically the \texttt{loss$\_$rate$\_$M} routine, to understand how the black hole's emission changes. \texttt{BlackHawk}'s \texttt{loss$\_$rate$\_$M} routine works by first computing the Page factor $\alpha(M)$ and storing the results in a table; using that table, it then computes the mass loss rate from Eq.~\eqref{dmdt}. For our purposes, we have added the additional powers $M$ and $M_p$ corresponding to a chosen $n_{MB}$ value. Once again, we emphasize that this is an order of magnitude estimate and its usefulness lies in its ability to predict distinguishable features of the emission spectra based on the new functional dependence, rather than precise predictions for the evaporation lightcurve and spectrum.

Using \texttt{BlackHawk}, in the standard emission picture with $n_{MB} = 0$, a black hole of $M_{\rm BH}^{(0)} = 5 \times 10^{14}$ g will have a lifetime of $1.126 \times 10^{17}$ s. To match this lifetime within an order of magnitude, we find (using the modified \texttt{BlackHawk loss$\_$rate$\_$M routine}) that for $n_{MB} = 1$ a black hole of mass $M_{\rm BH}^{(1)} = 1 \times 10^{7}$ g will give the Universe's lifetime within an order of magnitude. Similarly for $n_{MB} = 2$, \texttt{BlackHawk} finds a mass of $M_{\rm BH}^{(2)} = 5 \times 10^3$ g to match the Universe's lifetime within an order of magnitude. The precise numerical values we obtain are slightly different from the approximate result of Eq.~\eqref{mass_Mb}, as expected. 

Recall from Sec.~\ref{sec:theory} that the black hole's photon emission rate peaks when the parameter $x = E/T \sim 6$. Here however, the temperature evolution of the black hole follows a different power law relationship with time than the one in Eq. \eqref{timevtemp}. Using Eq.~\eqref{temp} and \eqref{mb_mass_loss}, we find
\begin{align}\label{temp MB}
    \tau^{(n_{MB})} \sim T^{-(3+2n_{MB})}\,,\quad\quad \tau^{(n_{MB})}_{peak} \sim E^{-(3+2n_{MB})}
\end{align}

Thus, the time of peak emission for photons of energy $E$ happens {\it earlier} for a memory burdened black hole than for a standard black hole. In Section \ref{sec:results} we discuss the implications of this relation on the photon lightcurves.

\section{Results}\label{sec:results}
This section presents our results on observational prospects to detect and measure the presence of new, secluded degrees of freedom and of memory burden effects using gamma-ray observatories. We first introduce, in the next section, the detectors we consider in our analysis, we then show  our results both for the case of additional degrees of freedom (sec.~\ref{sec:signals}), and for memory burden effects (sec.~\ref{subsec:memory burden})
\subsection{Detectors and Additional Dark Degrees of Freedom}\label{sec:signals}
As mentioned above, we consider two different, complementary gamma-ray telescopes, operating in the following energy ranges:
\begin{itemize}
    \item HAWC: $100 \,\text{GeV} <  E < 10^5 \,\text{GeV} $;
    \item Fermi-LAT: $0.1\,\text{GeV} < E  < 300\,\text{GeV} $.
\end{itemize}

\begin{figure}[t] 
\centering
\mbox{\hspace{-1cm} \includegraphics[width=0.55\textwidth]{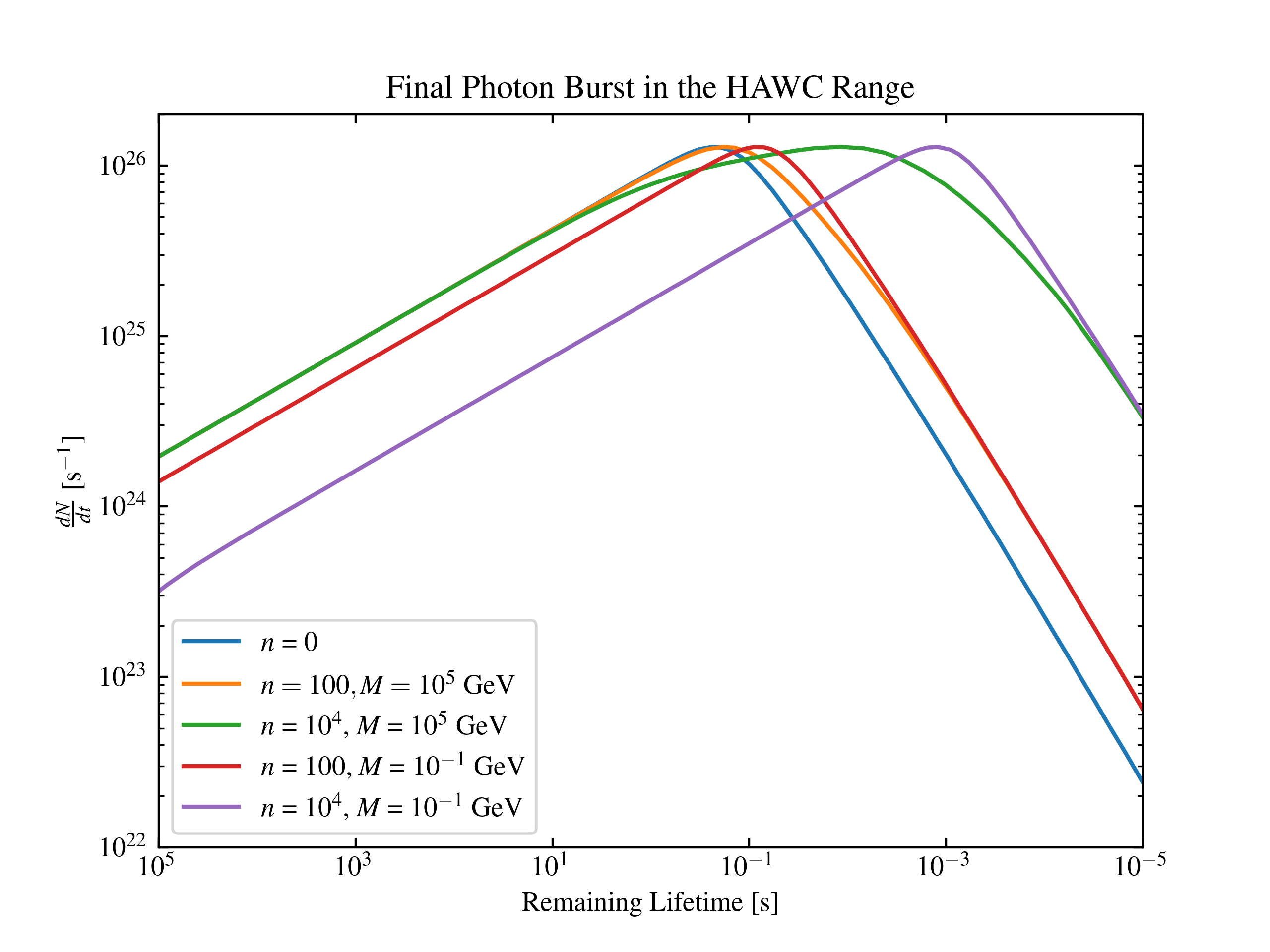}\hspace{-0.5cm} \includegraphics[width=0.55\textwidth]{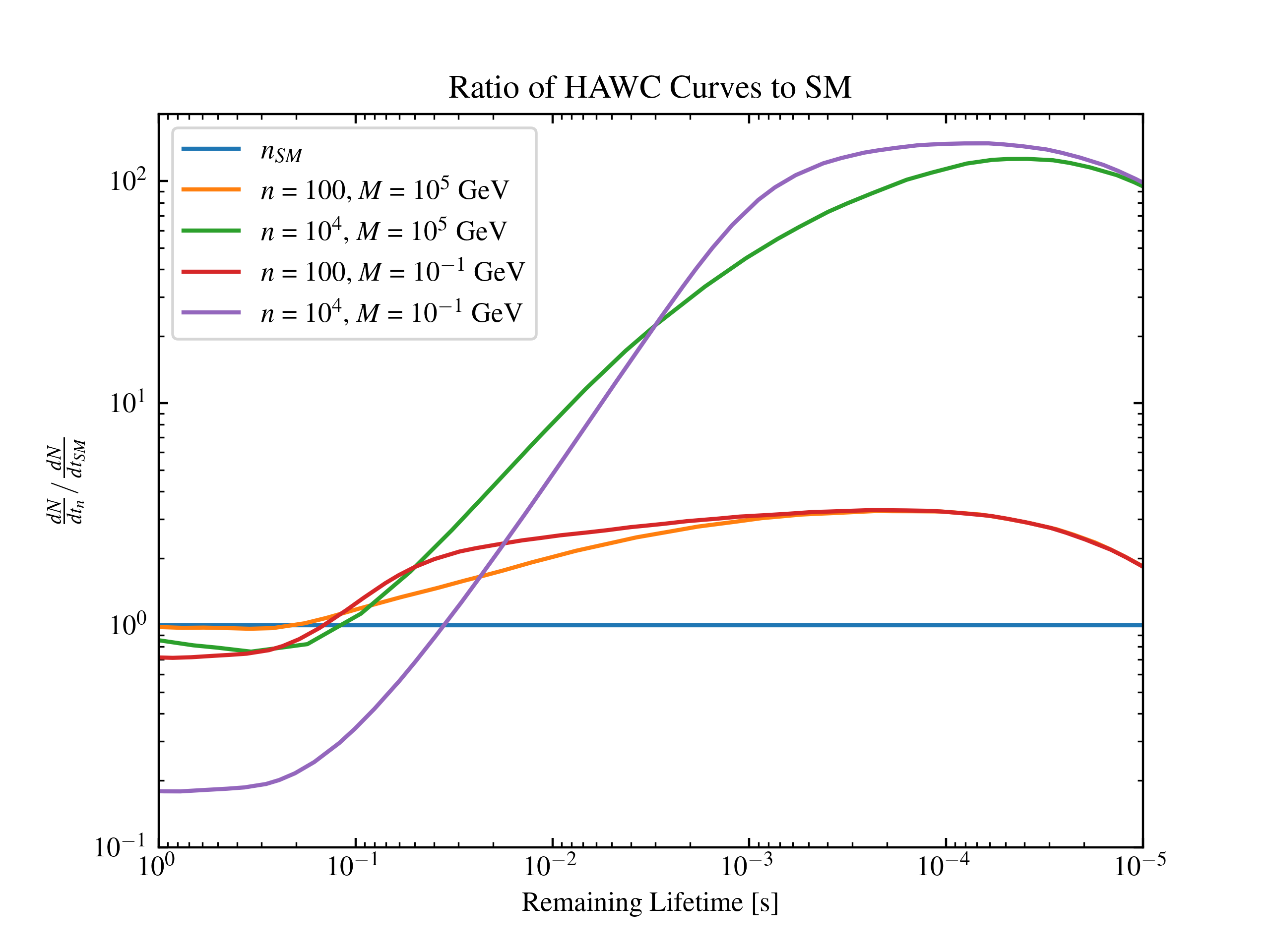}}
\caption{Final photon burst  of black hole evaporation integrated over the  HAWC energy range ($100 \,\text{GeV} <  E < 10^5 \,\text{GeV} $) with dark sectors at $M = 10^5$ GeV, $M = 10^{-1}$  GeV, and number of degrees of freedom $n=10^2,\ 10^4$; right: fluxes normalized to the Standard-Model-only case.}
\label{fig: HAWC Burst}
\end{figure}
\begin{figure}[t]
\centering
\mbox{\hspace{-1cm} \includegraphics[width=0.55\textwidth]{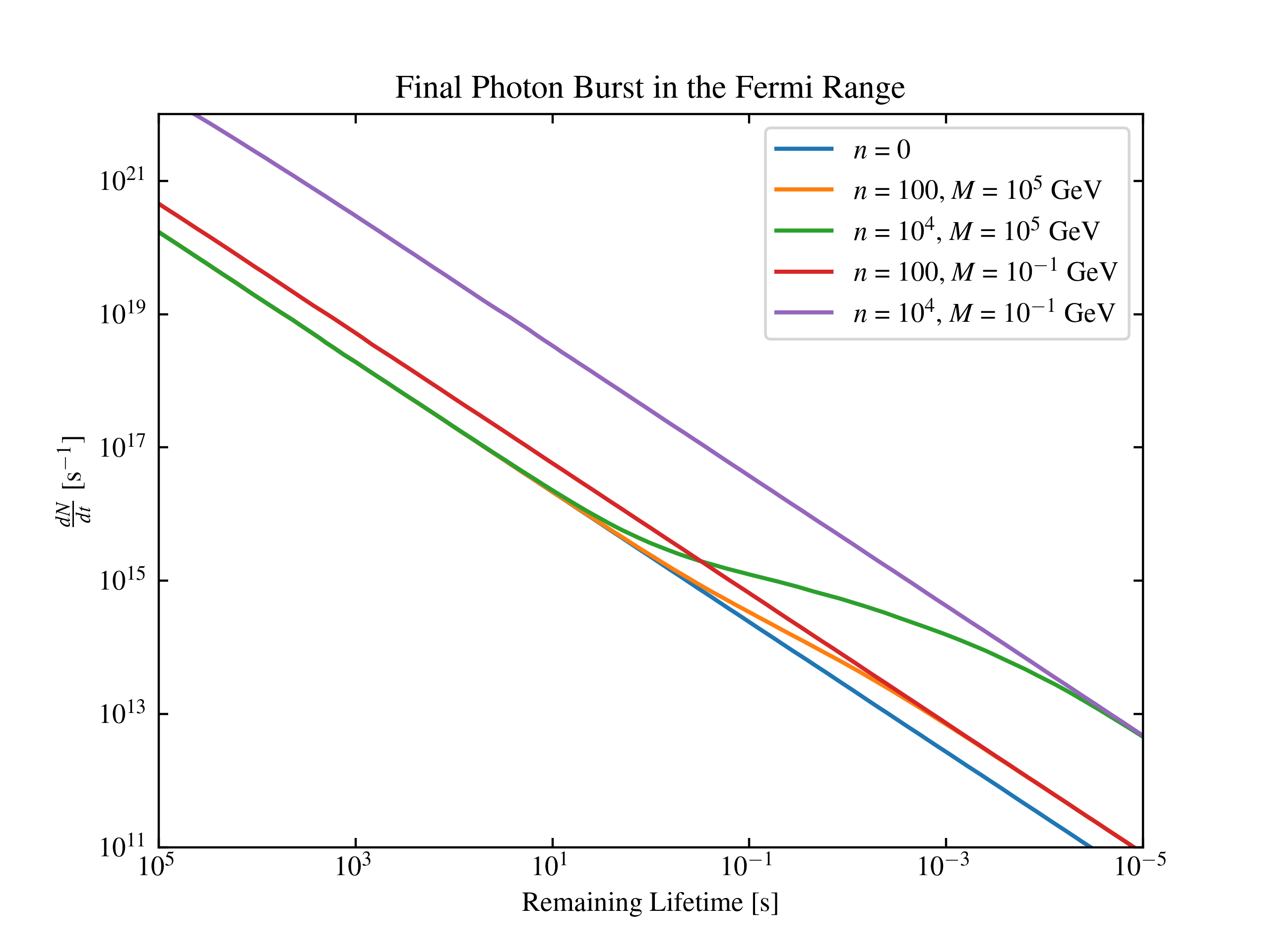}\hspace{-0.5cm} \includegraphics[width=0.55\textwidth]{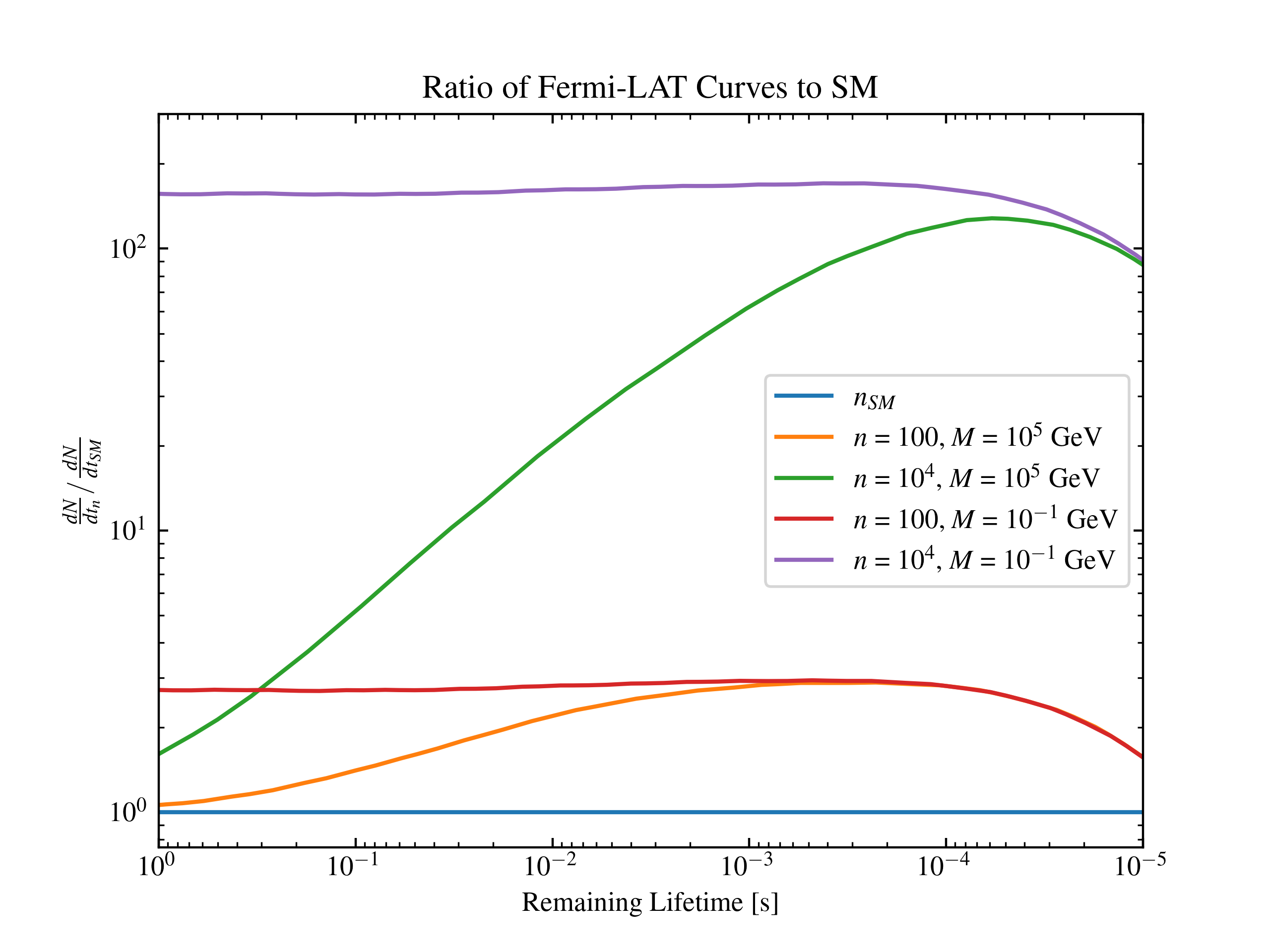}}
    \caption{Final photon burst of black hole evaporation integrated over the Fermi-LAT energy range ($0.1\,\text{GeV} < E  < 300\,\text{GeV} $) with dark sectors at $M = 10^5$ GeV, $M = 10^{-1} $ GeV, and number of degrees of freedom $n=10^2,\ 10^4$; right: fluxes normalized to the Standard-Model-only case.}
    \label{fig: Fermi Burst}
\end{figure}

We examine here models with additional degrees of freedom at common mass scales $M = 10^5 $ GeV and $M = 10^{-1} $ GeV. The first mass scale is indicative of degrees of freedom slightly beyond the current reach of present particle colliders, and the second could be associated with scenarios that include a  light dark matter candidate. 

We show   in figures \ref{fig: HAWC Burst} and \ref{fig: Fermi Burst}, for HAWC and Fermi-LAT respectively, the lightcurve of the final photon burst  corresponding to ($M,n$)=($10^5$ GeV, 100), ($10^5$ GeV, 10$^4$), ($10^{-1}$ GeV, 100), ($10^{-1}$ GeV, 10$^4$). The right panels are the lightcurves normalized to the $n=0$ case, i.e. the Standard Model expectation.

We now analyze of the asymptotic behaviors of the differential photon flux $dN/dt$ can be understood as appropriate limits of the relevant greybody factors, as a function of the maximal detector energy $E_{\rm max}$; specifically, in the geometric optics approximation, valid in the limit when the photon energy is much larger than the black hole's temperature, the greybody factor $\Gamma_H\sim M^2 E^2\sim E^2/T^2$, thus
\begin{equation}
    \left(\frac{dN}{dt}\right)_{E_{\rm max}\gg T}\sim\int \frac{E^2}{T^2}e^{-E/T}dE=T\int x^2 e^{-x}dx \sim T
\end{equation}
if the maximal energy is well above the black hole temperature $T$, i.e. $T\ll E_{\rm max}$, which, in the plots, corresponds to large $\tau$ (left part of the plot). 

If instead $E_{\rm max}\ll T$ we must consider the upper limit of integration; in addition, the low-energy grebody factor goes \cite{Page_1976_I} as $\Gamma_L\sim (ME)^{2s+1}$, where $s$ is the emitted particle spin, here photons, thus $s=1$; calling $x_{\rm max}\equiv E_{\rm max}/T$ we have
\begin{equation}
    \frac{dN}{dt}\sim T\int_0^{x_{\rm max}} x^3 e^{-x}dx =T(6+e^{-x_{\rm max}}(-6-x_{\rm max}(6+x_{\rm max}(3+x_{\rm max}))))\simeq T\frac{x^4_{\rm max}}{4}+{\cal O}(x_{\rm max}^5).
\end{equation}
where in the expression above we expanded to leading order in the small $x_{\rm max}$ limit; we thus finally find 
\begin{equation}
    \left(\frac{dN}{dt}\right)_{E_{\rm max}\ll T}\sim T\ x_{\rm max}^4\sim \frac{E_{\rm max}^4}{T^3}\sim T^{-3}\sim (\tau^{-1/3})^{-3}=\tau.
\end{equation}

The notable point the figures illustrate is that different dark sectors produce peaks at markedly different remaining lifetimes; specifically, the peaks emerge from the interplay of the lightcurve and of the energy range over which the experiment is sensitive to: when the black hole temperature exceeds the maximal energy to which the detector is sensitive, the photon count dramatically drops as $T^{-3}$; since the additional degrees of freedom affect the time-temperature relation, as discussed in detail above, this, in turn, induces a measurable change in the peak time/temperature.

Fig.~\ref{fig: HAWC Burst} and \ref{fig: Fermi Burst}, in particular, illustrate clearly how a larger number of degrees of freedom induces a peak at {\em smaller} remaining lifetime, if the mass-scale is within the energy range of the detector; additionally, a larger mass shifts the peak to larger remaining times, as the critical time $\tau_{\rm crit}$ is larger (see Tab.~\ref{tab:tcrit},  Fig.~\ref{fig:taucrit}, and Eq.~\eqref{critical time}).


\begin{figure}[t]
    \centering
\mbox{\hspace{-1cm} \includegraphics[width=0.55\textwidth]{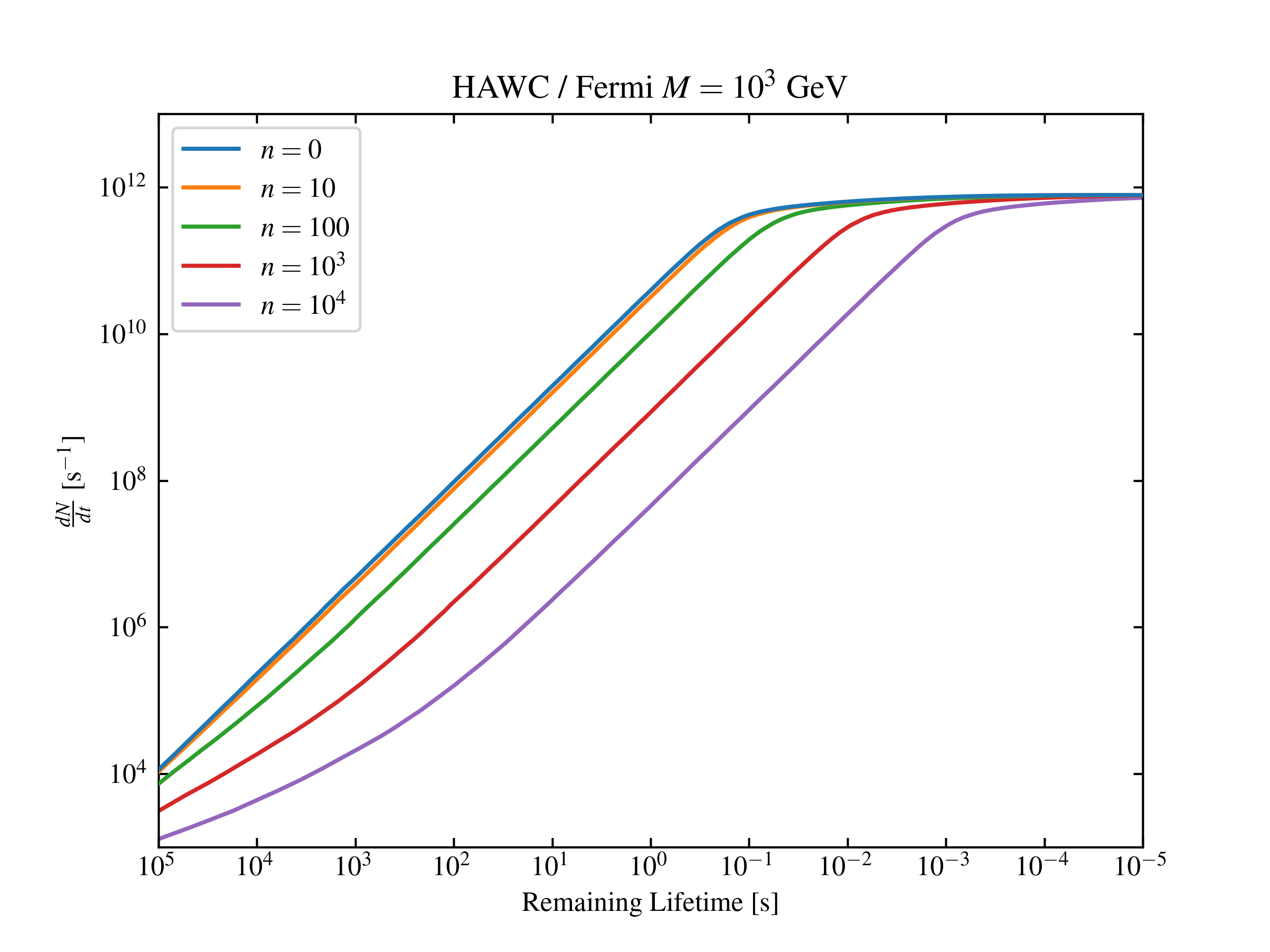}\hspace{-0.5cm} \includegraphics[width=0.55\textwidth]{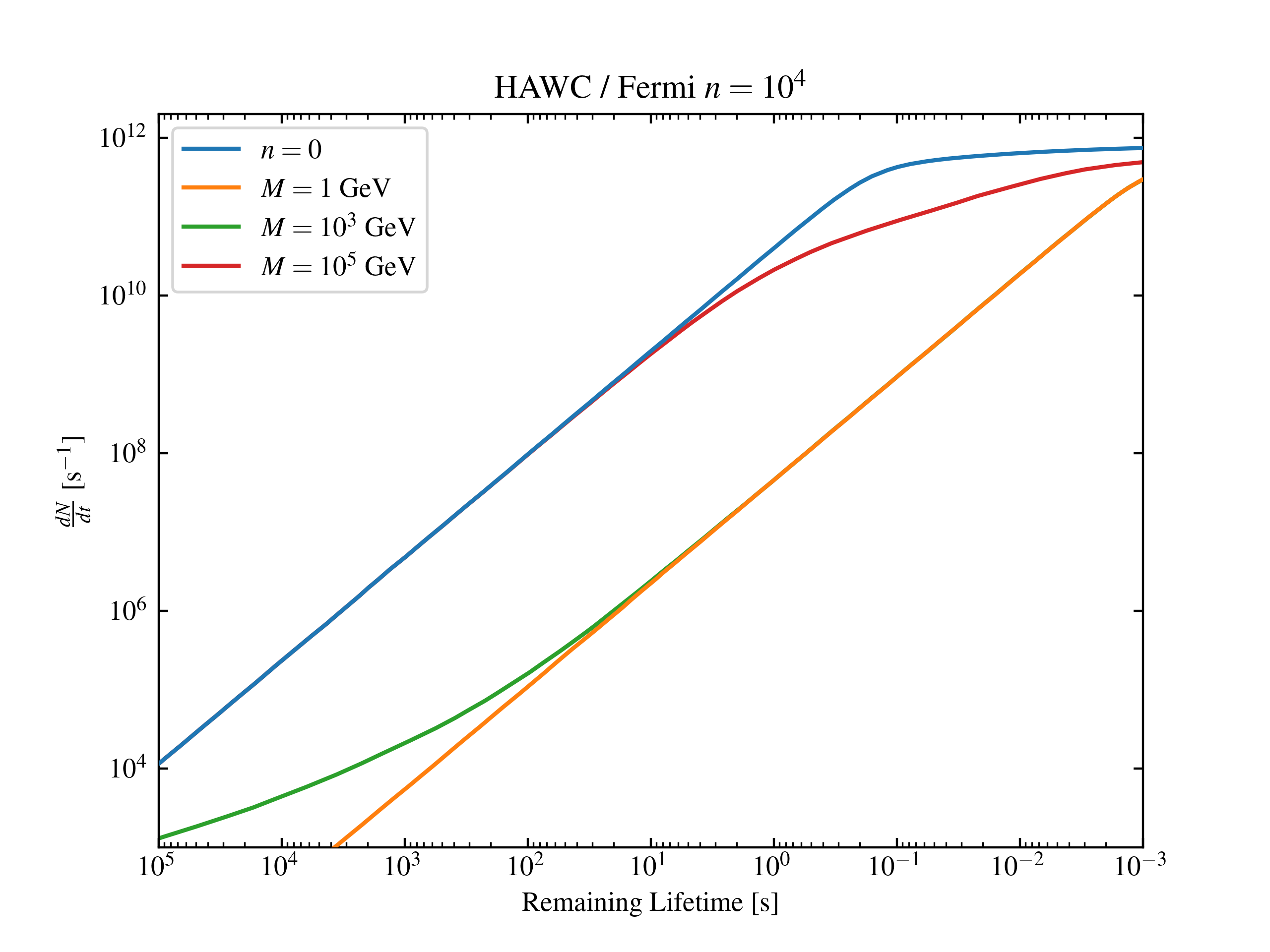}}
    \caption{Ratio of integrated photon fluxes in the HAWC over Fermi energy ranges, for varying mass scales and degrees of freedom.}\label{fig:HAWCtoFermi}
\end{figure}

If the mass scale falls outside the detector's mass range, as is the case for Fermi-LAT for $M=10^5$ GeV,  only a tail of the faster evaporation process will appear in the detector, if the number of additional degrees of freedom is large enough (see Fig.~\ref{fig: Fermi Burst}).

It also proves useful to compare the {\it ratio} of photon fluxes in each detector range. In figure \ref{fig:HAWCtoFermi} we show the ratio between HAWC and Fermi-LAT for an exploding black hole with dark sector $M = 10^3$ GeV and various numbers of degrees of freedom $n=10,\ 10^2,\ 10^3\ {\rm and}\ 10^4$ (left) and for $n=10^4$ and various mass scales $M=10^{-1},\ 10^3$ and $10^5$ GeV (right). The ratio plateaus at remaining lifetimes that correlate with the number of degrees of freedom, as expected from the discussion above of how $\tau_{\rm crit}$ scales with $n$. 

The left panel of Fig.~\ref{fig:HAWCtoFermi}, where we fix the new physics mass scale at $M=10^3$ GeV and vary $n$, shows how the ratio of the photon flux at HAWC relative to that at Fermi-LAT, as a function of time, is a diagnostic of the number of dark degrees of freedom, especially if the mass scale lies close to the energy range at which both telescopes are sensitive to. The right panel, where $n=1e4$ and we change $M$, shows how there is a relative small, but potentially measurable change to the HAWC/Fermi ratio, especially for mass scales within and beyond the Fermi-LAT energy range, with a distinct slope, which could be used to extract information on $M$. 

In detail, given the scaling of the critical temperature with $n$ and $M$ in Eq.~\eqref{critical time} above, and the fact that the peak corresponds to $T\sim E_{\rm max}$, we find that the position of the peak scales as $n^{-1/3} \ln n$; the scaling with $M$ is reflective of the relation defining the peak as the intercept of the asymptotic $\sim T$ ($T\ll E_{\rm crit}$) and $\sim T^{-3}$ ($T\gg E_{\rm crit}$), and we find that the position of the peak, as a result $\sim M^{-1/4}$, hence the small effect we mentioned above.

\begin{figure}[t]
    \centering
    \includegraphics[width=0.75\textwidth]{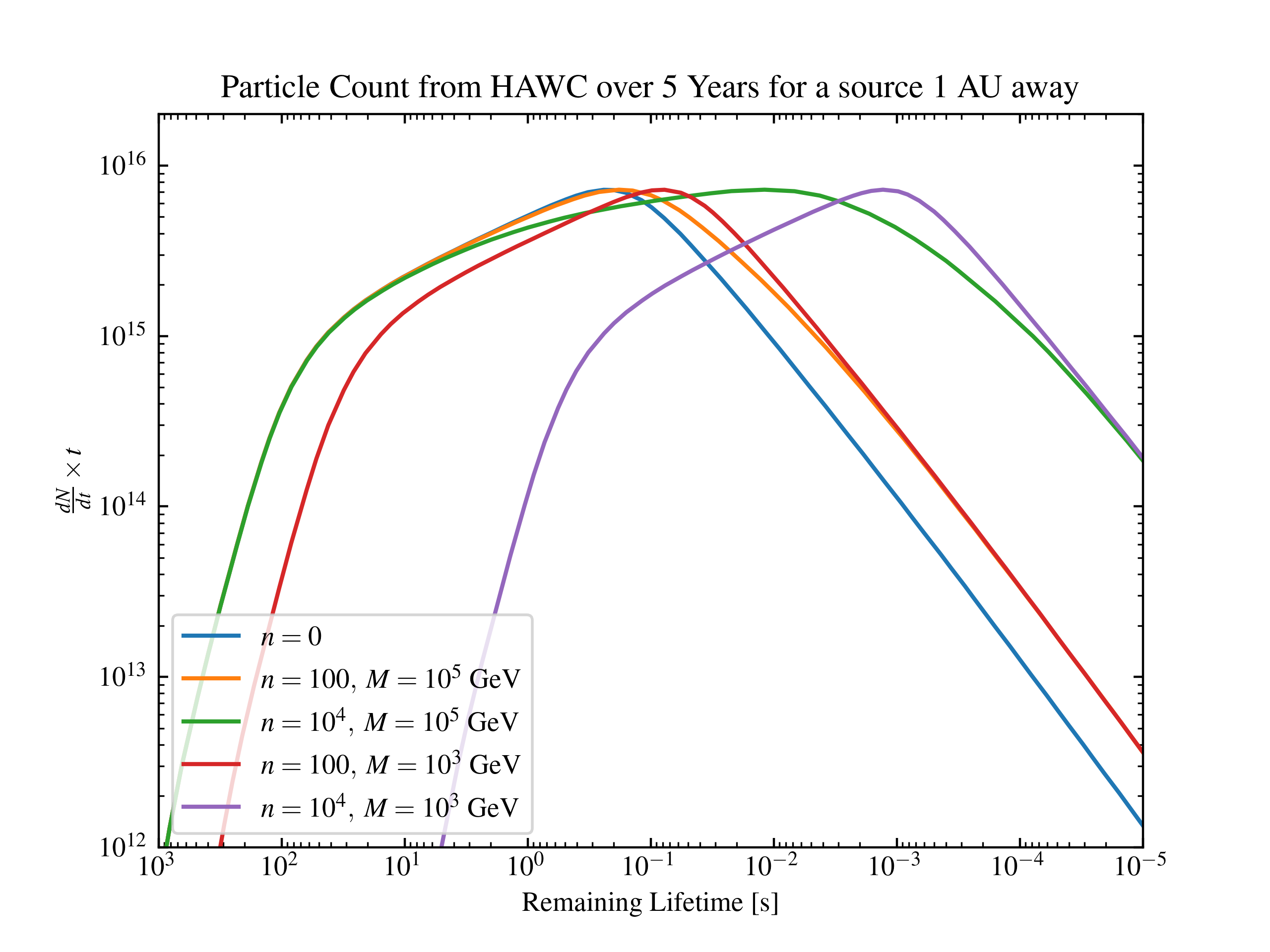}
    \caption{The quantity $\tau\ dN/dt$ HAWC would observe during the last $\sim$ hour of an evaporating black hole's lifetime from a distance of 1 AU.}
    \label{fig: particlecounthawc}
\end{figure}

Figure \ref{fig: particlecounthawc} shows the product of the photon flux times the remaining lifetime, for a putative source 1 AU away; this quantity is a proxy for the time at which the largest photon count is expected. Notice that the number of degrees of freedom $n$ is correlated exactly with the asymptote at small remaining lifetimes, while the mass scale with the location of the peak emission --- as before, larger mass scales peak at smaller remaining lifetimes. This is therefore an additional diagnostic for both $n$ and $M$. The photon accumulation rate is thus very slow at early times, it peaks in the $n$ and $M$ model dependent way discussed, and it decreases, because of the short time scale, near the end of evaporation.

\begin{figure}[t]
    \includegraphics[width=0.75\textwidth]{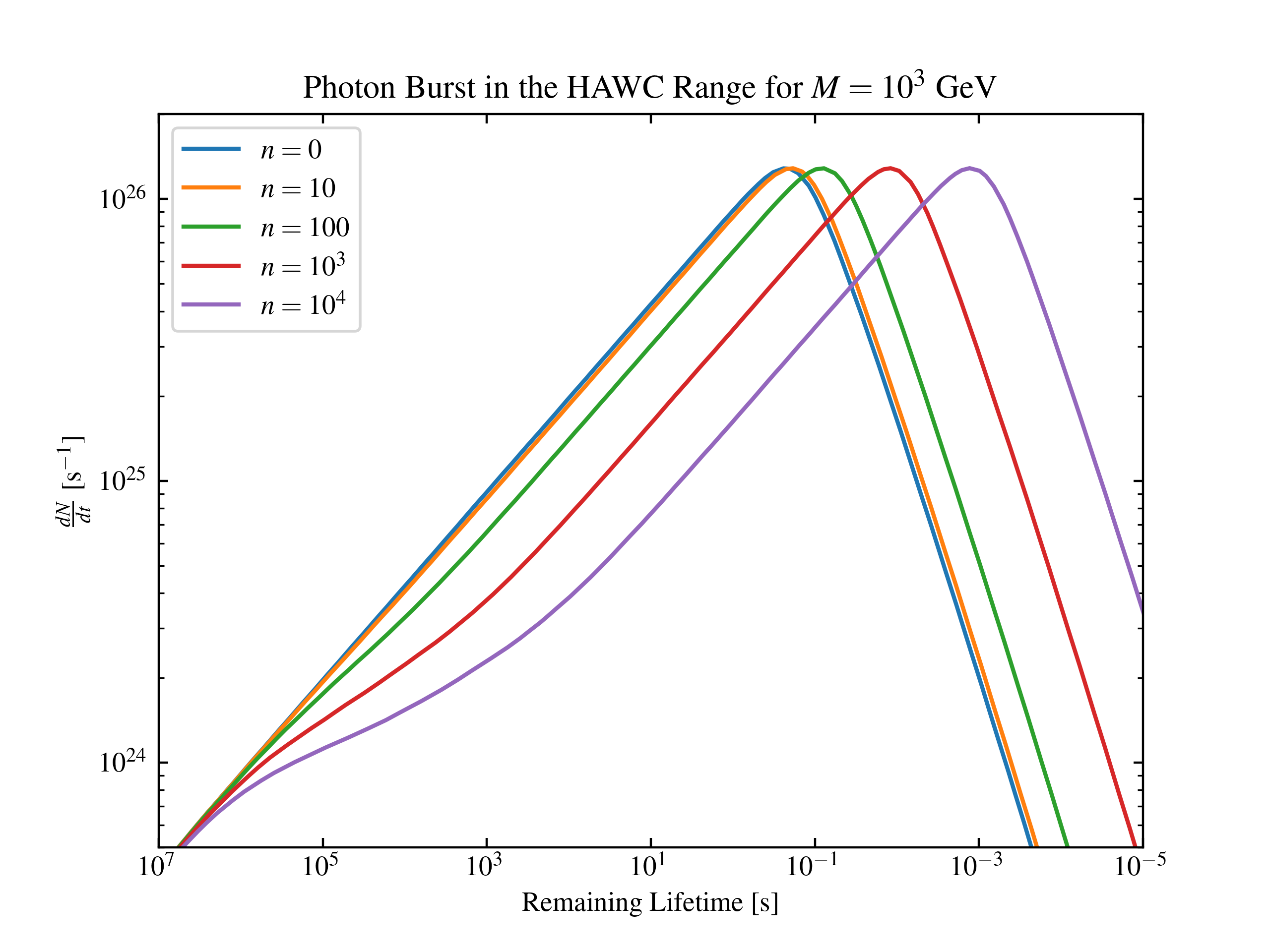}
    \caption{Final photon burst, integrated over the HAWC energy range, for an evaporating black hole with dark sectors of various size $n$, at $M = 10^3$ GeV.}
    \label{fig: HAWC m1e3}
\end{figure}

Our final figure \ref{fig: HAWC m1e3} shows the differential photon rate as a function of $n$ for a value of $M=10^3$ GeV within the energy range of the observatory, HAWC in this example. Again, we reproduce the asymptotic, large $n\gg n_{\rm SM}$ scaling with $n$ of the the differential flux $dN/dt\sim n^{-1/3} \ln n$, thus conclusively offering a diagnostic for $n$.




While in principle we could consider additional mass scales representing different models of light dark matter -- say $M = 10^{-4} $ GeV to $M = 10^{-20} $ GeV, they would have the same exact emission profiles as the $M = 10^{-1}$ GeV curve of corresponding $n$ (see table \ref{tab:tcrit}).

This means that, in principle, one could put bounds on the mass scale of dark, secluded degrees of freedom detectable via a black hole burst using Eq. \eqref{critical time}. For example, if the telescope can only observe for say 10 years, then the lowest mass scale that one could probe would be $M \sim 100$ GeV for $n = 100$ and $M \sim 250$ GeV for $n = 10^4$. Similarly, the highest mass scale observable can be calculated. If the telescope cannot resolve anything past $10^{-3}$ s, then the highest observable mass scale would be $M \sim 1.7 \times 10^{6}$ for $n = 100$, and $M \sim 3.7 \times 10^{6}$ for $n = 10^4$.

\subsection{Memory Burden}\label{subsec:memory burden}
\begin{figure}[t]
    \includegraphics[width=0.75\textwidth]{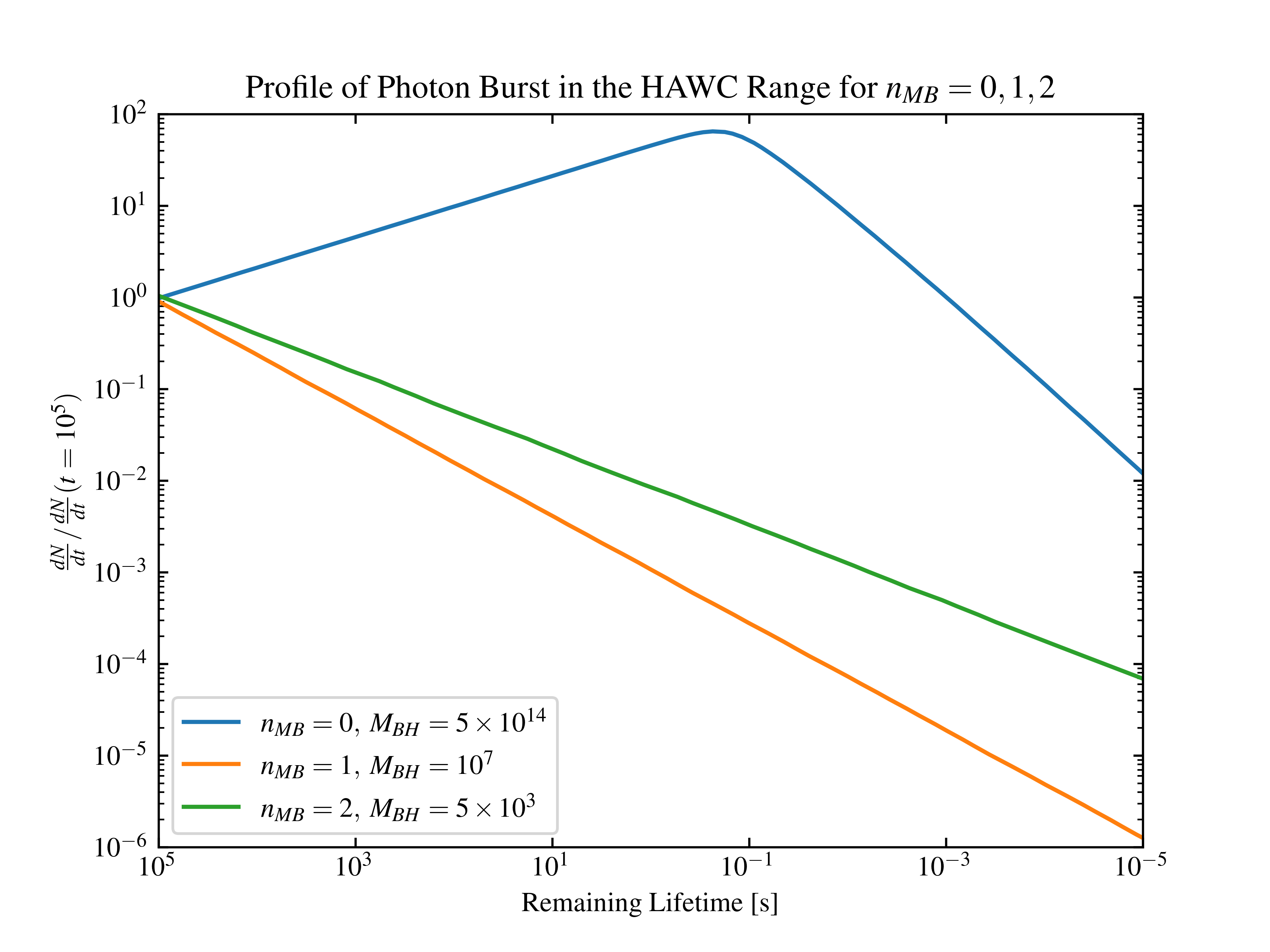}
    \caption{Black hole emission rate integrated over the HAWC energy range for black holes exploding today with $n_{MB} = 0, 1, 2$, normalized at $t = 10^5$ sec.}
    \label{fig: MB emission}
\end{figure}
\begin{figure}[t]
    \includegraphics[width=0.75\textwidth]{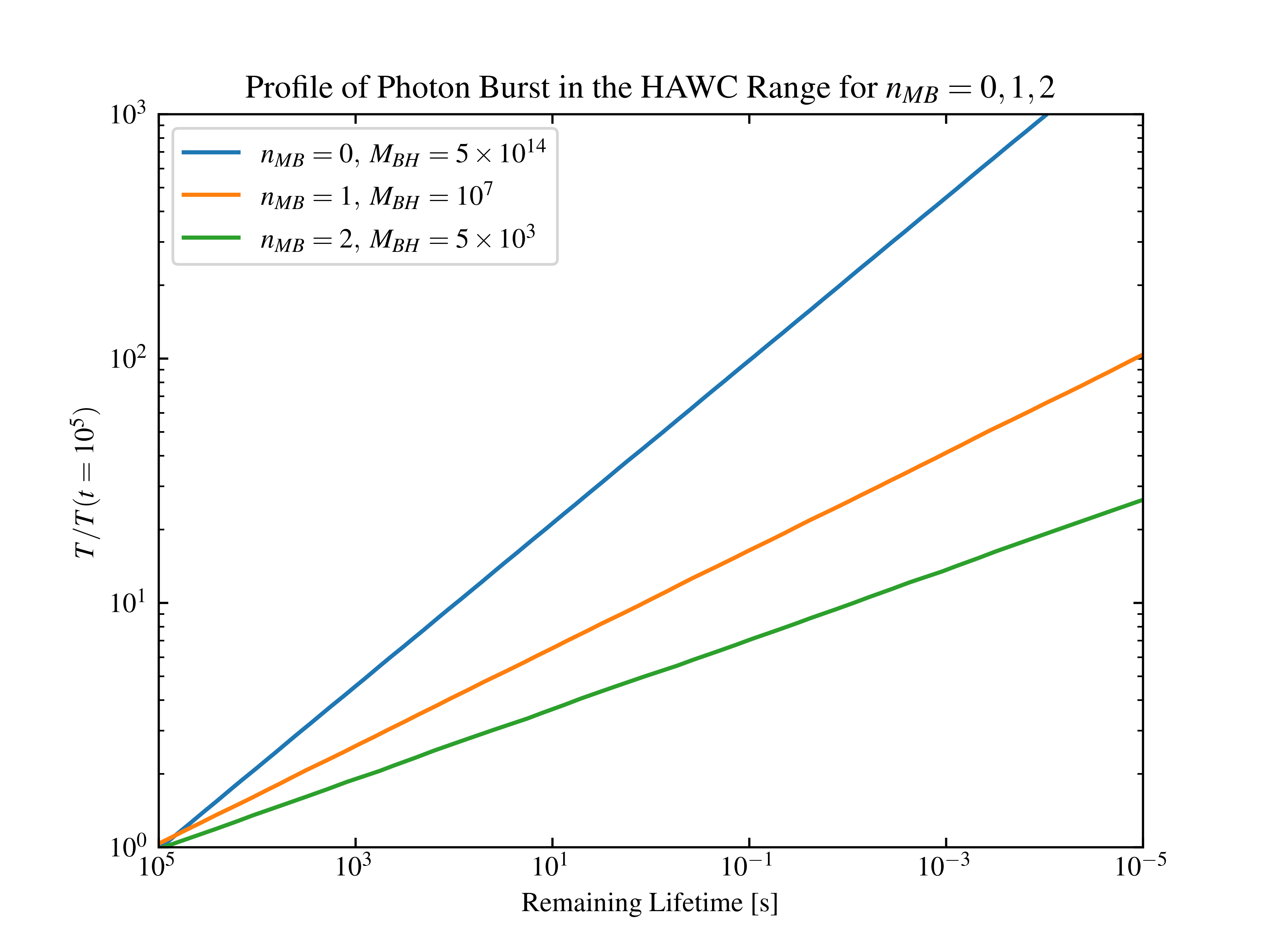}
    \caption{Black hole remaining lifetime-temperature relation in the HAWC range for black holes exploding today with memory burden coefficients $n_{MB} = 0, 1, 2$, normalized at $t = 10^5$ sec.}
    \label{fig: MB temp}
\end{figure}

Memory burden effects slow down the rate of evaporation of black holes. This means the black holes exploding today would be {\it much lighter} than in the standard framework. Furthermore, as Eq. \eqref{temp MB} shows, their emission {\it rate} is also altered. 

In Fig.~\ref{fig: MB emission}, where we plot the evaporation rate normalized to a time 10$^5$ seconds to complete evaporation, one can see that the emission  rate of black holes burdened by memory is significantly suppressed. Specifically, the asymptotic power-law, at temperatures {\em larger} than the maximal energy of the observatory (here, 10$^5$ GeV for HAWC) reads
\begin{equation}
\label{eq:tempTmb}
    \frac{dN}{dt}\sim T^{-3}\sim \tau^{\frac{3}{3+2n_{MB}}},
\end{equation}
a behavior asymptotically reproduced in the figure at small-enough times (corresponding to large temperatures).

The final plot exhibits the memory burden time-temperature relation as a function of the remaining time $\tau$, and reflects, once again, the behavior anticipated in Eq.~\eqref{eq:tempTmb}.

Memory burden effects can be directly measured by  taking ratios of photon counts at large and small remaining times to expiration, say in an early-time (large $\tau$) bin $(\Delta t)_{\rm early}$, with a total number of collected photons $N_{\gamma,{\rm early}}$, and in a late time (small $\tau$) bin $(\Delta t)_{\rm late}$, with a total number of collected photons $N_{\gamma,{\rm late}}$ extracting $n_{MB}$ from 
\begin{equation}
    \frac{\ln\left(\frac{N_{\gamma,{\rm early}}}{N_{\gamma,{\rm late}}}\right)}{\ln\left(\frac{(\Delta t)_{\rm early}}{(\Delta t)_{\rm early}}\right)}=R=\frac{3}{3+2n_{MB}},
\end{equation}
giving
\begin{equation}
    n_{MB}=\frac{3-3R}{R}.
\end{equation}
For instance, for $n_{MB}=2$ we find that
\begin{equation}
    R\simeq \frac{\ln(1./10^{-6})}{\ln(10^5/10^{-5})}=\frac{3}{5},\quad n_{MB}=\frac{3}{3+2n_{MB}}=\left(3-3\frac{3}{5}\right)\frac{5}{3}=2.
\end{equation}


\section{Discussion and Conclusion}\label{sec:discussion}
The final phase of black hole evaporation is a remarkable window on potentially utterly secluded dark sectors, and, more generally, on new physics: the evaporation rate of the black hole is affected both by the presence of any additional degrees of freedom to the Standard Model of elementary particles and interactions, and on quantum gravity effects affecting Hawking evaporation when temperatures approach the Planck scale. Interestingly, the final phase of the evaporation of a black hole is potentially observable if a population of black holes exists with masses significantly lighter than those of stellar origin. The existence of such a population is well motivated by a number of formation mechanisms, and could explain, in part or in totality, the nature of the cosmological dark matter \cite{Carr:2021bzv}.

In the case of additional degrees of freedom, a simple parametrization requires specifying (i) the mass scale $M$ of such degrees of freedom and (ii) the number of such degrees of freedom $n$. In the present study we have focused on observational probes of both quantities, and obtained new analytical insights on how to derive the values of these quantities from observations.

Specifically, we showed that the time-temperature relation, measurable from the lightcurve of the evaporation process, contains a feature at a critical time-to-expiration $\tau_{\rm crit}$ which depends on $M$ and $n$ as
\begin{equation}
    \tau_{\rm crit}(n,M) \sim\frac{\ln(n)}{M};
\end{equation}

Secondly, we showed that the (integrated) photon flux expected to be collected at an observatory with a maximal energy sensitivity $E_{\rm max}$ has the low- and high-temperature (respectively long and short remaining lifetime) asymptotic behaviors
\begin{equation}\label{eq:asym}
    \frac{dN}{dt}\sim T\quad (T\ll E_{\rm max});\ \quad  \frac{dN}{dt}\sim T^{-3}\quad (T\gg E_{\rm max});
\end{equation}
the corresponding asymptotic behaviors in the remaining lifetime $\tau$ follow from the relation \mbox{$T\sim \tau^{-1/3}$};

Thirdly, we showed that the position of the peak resulting from the asymptotic structure in Eq.~\eqref{eq:asym} results in a dependence on $n$ which reads $$\tau_{\rm peak}(n)\sim (n^{-1/3})(\ln n)/(M^{1/4}).$$ The peak's position is also reflected in the ratio of the HAWC-to-Fermi-LAT lightcurves (Fig.~\ref{fig:HAWCtoFermi}, right panel). 
Were the three quantities above measured directly, one can thus obtain a relatively conclusive measurement of both $n$ and $M$.

We also discussed both analytically and numerically the precise scaling of the lightcurve with the memory burden scaling of black hole lifetime with entropy/mass of the black hole in the presence of memory burden effects characterized by an index $n_{MB}$. We showed that one can reconstruct the crucial quantity $n_{MB}$ from measurements of the ratio of photon flux at different lifetimes, as 
\begin{equation}
    n_{MB}=\frac{3-3R}{R},\quad {\rm where}\quad R=\frac{\ln\left(\frac{N_{\gamma,{\rm early}}}{N_{\gamma,{\rm late}}}\right)}{\ln\left(\frac{(\Delta t)_{\rm early}}{(\Delta t)_{\rm early}}\right)}.
\end{equation}

Concluding, a measurement of the last instants of the evaporation of a light black hole provides a window on, and a concrete and viable way to measure, the mass-scale and number of additional secluded, dark degrees of freedom, as well as possible quantum gravity effects that could drive a slower evaporation time-scale near the final stages of evaporation.

\section*{Acknowledgments} \label{sec:acknowledgements}
We are grateful to Jos\'e Ramon Espinosa for suggesting the exploration of memory burden effects in the context of black hole explosions. This work is partly supported by the U.S.\ Department of Energy grant number de-sc0010107 (SP).

\bibliographystyle{bibi} 
\bibliography{bib}


\end{document}